\documentclass[prl,twocolumn,amsmath,amssymb,longbibliography,superscriptaddress]{revtex4-1}

\pdfoutput=1

\usepackage[pdftex]{graphicx}
\usepackage[pdftex,colorlinks=true,urlcolor=blue,linkcolor=black]{hyperref} 

\usepackage{graphicx}
\usepackage{dcolumn}
\usepackage{bm}
\usepackage{datetime}
\usepackage{color}

\usepackage{enumitem}
\setitemize{leftmargin=*} 

\usepackage{ulem}				

\newcommand{\rmi}{{\rm i}}

\begin{document}

\title{Observation of the Berezinskii-Kosterlitz-Thouless Phase Transition\\ in an Ultracold Fermi Gas}

\author{P. A. Murthy}
\email{murthy@physi.uni-heidelberg.de}
\affiliation{Physikalisches Institut, Ruprecht-Karls-Universit\"at Heidelberg, 69120 Heidelberg, Germany}

\author{I. Boettcher}
\email{i.boettcher@thphys.uni-heidelberg.de}
\affiliation{Institut f\"ur Theoretische Physik, Ruprecht-Karls-Universit\"at Heidelberg, 69120 Heidelberg, Germany}

\author{L. Bayha}
\affiliation{Physikalisches Institut, Ruprecht-Karls-Universit\"at Heidelberg, 69120 Heidelberg, Germany}

\author{M. Holzmann}
\affiliation{LPTMC, UMR 7600 of CNRS, Universit\'{e} Pierre et Marie Curie, Paris, France}
\affiliation{Universit\'{e} Grenoble Alpes, CNRS, LPMMC, UMR 5493, F-38042 Grenoble, France}

\author{D. Kedar}
\affiliation{Physikalisches Institut, Ruprecht-Karls-Universit\"at Heidelberg, 69120 Heidelberg, Germany}

\author{M. Neidig}
\affiliation{Physikalisches Institut, Ruprecht-Karls-Universit\"at Heidelberg, 69120 Heidelberg, Germany}

\author{M. G. Ries}
\affiliation{Physikalisches Institut, Ruprecht-Karls-Universit\"at Heidelberg, 69120 Heidelberg, Germany}

\author{A. N. Wenz}
\affiliation{Physikalisches Institut, Ruprecht-Karls-Universit\"at Heidelberg, 69120 Heidelberg, Germany}

\author{G. Z\"urn}
\affiliation{Physikalisches Institut, Ruprecht-Karls-Universit\"at Heidelberg, 69120 Heidelberg, Germany}

\author{S. Jochim}
\affiliation{Physikalisches Institut, Ruprecht-Karls-Universit\"at Heidelberg, 69120 Heidelberg, Germany}

\date{\today}

\begin{abstract}
We experimentally investigate the first-order correlation function of a trapped Fermi gas in the two-dimensional BEC-BCS crossover. We observe a transition to a low-temperature superfluid phase with algebraically decaying correlations. We show that the spatial coherence of the entire trapped system can be characterized by a single temperature-dependent exponent. We find the exponent at the transition to be constant over a wide range of interaction strengths across the crossover.
This suggests that the phase transitions in both the bosonic regime and the strongly interacting crossover regime are of Berezinskii-Kosterlitz-Thouless type and lie within the same universality class. On the bosonic side of the crossover, our data are well-described by the quantum Monte Carlo calculations for a Bose gas. In contrast, in the strongly interacting  regime, we observe a superfluid phase which is significantly influenced by the fermionic nature of the constituent particles. 
\end{abstract}

\pacs{Valid PACS appear here}
\maketitle

Long-range coherence is the hallmark of superfluidity and Bose-Einstein condensation \cite{Penrose1956, Yang1962}. The character of spatial coherence in a system and the properties of the corresponding phase transitions are fundamentally influenced by dimensionality. The two-dimensional case is particularly intriguing as for a homogeneous system, true long-range order cannot persist at any finite temperature due to the dominant role of phase fluctuations with large wavelengths \cite{Mermin1966, Hohenberg1967, Kadanoff1967}. Although this prevents Bose-Einstein condensation in 2D, a transition to a superfluid phase with quasi-long-range order can still occur, as pointed out by Berezinskii, Kosterlitz, and Thouless (BKT) \cite{Berezinskii1972, Kosterlitz1973, Kosterlitz1974}. A key prediction of this theory is the scale-invariant behavior of the first-order correlation function $g_1(r)$, which, in the low-temperature phase, decays algebraically according to $g_1(r)\propto\,r^{-\eta}$ for large separations $r$. Importantly, the BKT theory for homogeneous systems predicts a universal value of $\eta_{\rm c} = 1/4$ at the critical temperature, accompanied by a universal jump of the superfluid density \cite{Nelson1977}.

Several key signatures of BKT physics have been experimentally observed in a variety of systems such as exciton-polariton condensates \cite{Roumpos2012}, layered magnets \cite{Durr1989, Ballentine1990},  liquid $^4$He films \cite{Bishop1978}, and trapped Bose gases \cite{Hadzibabic2006, Clade2009, Tung2010,  Hung2011, Plisson2011,Desbuquois2012,choi2013}. Particularly in the context of superfluidity, the universal jump in the superfluid density was measured in thin films of liquid $^4$He \cite{Bishop1978}. More recently, in the pioneering interference experiment with a weakly interacting Bose gas \cite{Hadzibabic2006}, the emergence of quasi-long-range order and the proliferation of vortices were shown. 

There are still important aspects of superfluidity in two-dimensional systems that remain to be understood, which we aim to elucidate in this work with ultracold atoms. One question is whether the BKT phenomenology can also be extended to systems with nonuniform density. Indeed, if the microscopic symmetries are the same, the general physical picture involving phase fluctuations should be valid also for inhomogeneous systems. However, it is not known if algebraic order persists at all in the presence of inhomogeneity and particularly, whether the correlations in the whole system can still be characterized by a single exponent. Another fundamental issue that arises in the study of superfluidity is the pairing of fermions. While fermionic superfluidity has been extensively investigated in 3D systems \cite{Zwierlein2005, Osheroff1972, Leggett1976}, there are open experimental questions in the 2D context. In particular, what is the long-range behavior of spatial coherence of a 2D fermionic superfluid, and can it also be described in the BKT framework like its bosonic counterpart?

In this work, we probe the first-order correlation function $g_1(r)$ of a trapped Fermi gas in the two-dimensional BEC-BCS crossover regime \cite{Ries2015, Levinsen2015}. The correlation function is determined from a measurement of the in situ momentum distribution of the gas. We demonstrate that even in this inhomogeneous system, algebraic order persists in $g_1(r)$ below a critical temperature. Furthermore, a quantitative analysis of the scaling exponents across the crossover reveals the validity of the BKT theory also in the fermionic regime.

Our measurements are performed with a gas of $10^5$ $^6$Li atoms confined in a highly anisotropic potential. The axial and radial trapping frequencies are $\omega_z \approx 2\pi \times 5.5\,$kHz and $\omega_r \approx 2\pi \times 18\,$Hz, leading to an aspect ratio of approximately 300:1. Our experimental system and methodology have been described in detail in Ref.\,\cite{Ries2015}. We perform in situ imaging of the sample as a function of temperature and interaction strength. From the central density, we define the Fermi momentum $k_{\rm F}$ and Fermi temperature $T_{\rm F}$, which constitute the relevant scales in the system. As shown in Ref.\,\cite{Ries2015}, for our experimental parameters, all the relevant energy scales are smaller than the axial confinement energy $\hbar \omega_z$. Hence the system is in the quasi-2D regime. 

We tune the interparticle interactions using a Feshbach resonance located at 832 G. Using the 3D scattering length $a_{\rm 3D}$ \cite{Zurn2013}, the axial oscillator length $\ell_z$ \footnote{We define $\ell_z=\sqrt{\hbar/M\omega_z}$, where $M$ is twice the fermion mass.}, and the Fermi momentum, we construct the effective 2D scattering length $a_{\rm 2D}$ and crossover parameter $\ln(k_{\rm F}a_{\rm 2D})$ \cite{Levinsen2015}. For $\ln(k_{\rm F}a_{\rm 2D})\ll -1$ and $\ln(k_{\rm F}a_{\rm 2D}) \gg1$ we are in the bosonic and fermionic limit of the crossover, respectively. 

In addition to the measurements, we perform path-integral quantum Monte Carlo (QMC) computations of a Bose gas \cite{Holzmann2008, Holzmann2010} in a highly anisotropic 3D trap with parameters similar to those employed in the experiment. In the simulations, the bosons interact via the molecular scattering length $a_{\rm mol} = 0.6\,a_{\rm 3D}$ \cite{Petrov2004}. The relevant parameters that describe the system in terms of pointlike bosons are the effective bosonic coupling strength $\tilde{g} = \sqrt{8\pi} a_{\rm mol}/\ell_z$ and the condensation temperature of an ideal 2D Bose gas $T_{\rm BEC}^0 = \sqrt{6N} \frac{\hbar \omega_{r}}{\pi k_{\rm B}} \approx 140\,$nK, where $N$ is the number of particles. We use these bosonic parameters to compare our measurements to QMC at the lowest magnetic field values, where we have $\tilde{g} = 0.6, 1.07, 2.76, 7.75$ \cite{SOM}. From the QMC computations, we obtain the local density profile and the one-body density matrix $\rho_1(\textbf{x},\textbf{x}')=\langle \hat{\phi}^\dagger(\textbf{x})\hat{\phi}(\textbf{x}')\rangle$ for different interaction strengths and temperatures, where $\hat{\phi}(\textbf{x})$ is the bosonic field operator. 

The global off-diagonal correlations in the system are encoded in the momentum distribution of particles. To reliably measure the in-plane momentum distribution $\tilde{n}(\textbf{k})$ of our sample, we employ the matterwave focusing technique described in Refs.\,\cite{Shvarchuck2002, Tung2010, Murthy2014}, where the gas expands freely in the axial direction while being focused by a harmonic potential in the radial plane. After expansion for a quarter of the period of the focusing potential, the initial momentum distribution is mapped to the spatial density profile, which we then image. We combine this focusing method with a rapid magnetic field ramp into the weakly interacting regime. This rapid ramp technique -- along with the fast axial expansion due to the large anisotropy of the trap -- ensures that inter-particle collisions during the focusing do not cause significant distortions to the measured momentum distribution. From $\tilde{n}(\textbf{k})$, we extract the absolute temperature $T$ by means of a Boltzmann fit to the high-$\textbf{k}$ thermal region \footnote{The temperatures accessed in this work range between 40\,nK and 150\,nK}. 

\begin{figure} [ht!]
\includegraphics [width=8.5cm] {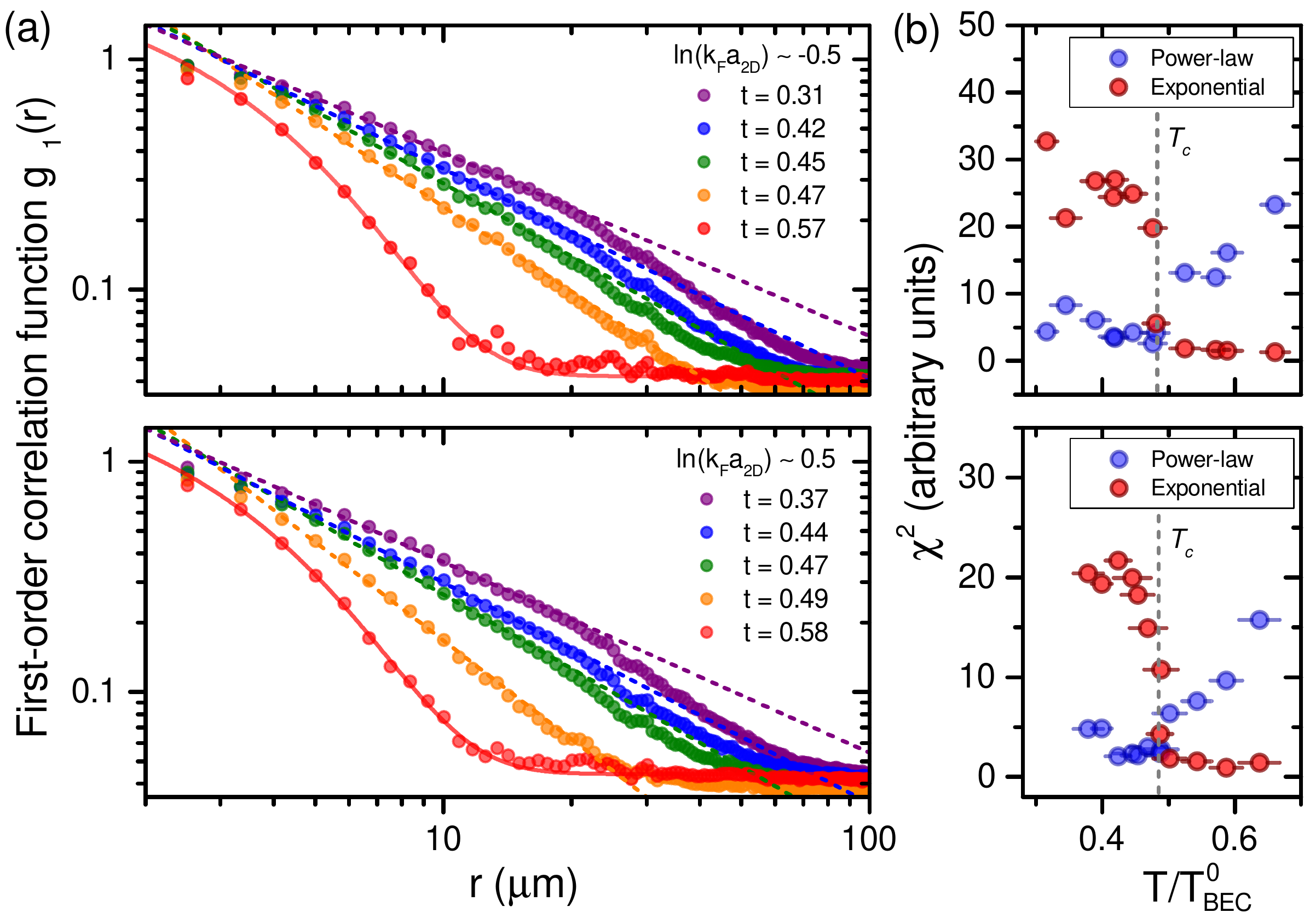}
\caption{First-order correlation function $g_1(r)$ for different temperatures at $\ln(k_{\rm F}a_{\rm 2D}) \simeq -0.5$ (upper left panel) and $\ln(k_{\rm F}a_{\rm 2D}) \simeq 0.5$ (lower left panel). The temperature scale used here is $t = T/T^0_{\rm BEC}$. At high temperatures correlations decay exponentially as expected for a gas in the normal phase. At low temperatures, we observe algebraic correlations ($g_1(r) \propto r^{-\eta(T)}$) with a temperature-dependent scaling exponent $\eta(T)$. This qualitative change of behavior is clearly visible in the $\chi^2$ for both exponential and algebraic fits (right panel), where a small value signals a good fit. In particular, this allows for an accurate determination of the transition temperature $T_{\rm c}$ (vertical dashed lines) \cite{SOM}.}
\label{fig:g1r}
\end{figure}

To quantitatively investigate the spatial coherence in our system, we determine the first-order correlation function $g_1(\textbf{r})$ by means of a 2D Fourier transform of the measured $\tilde{n}(\textbf{k})$. It is related to the one-body density matrix $\rho_1(\textbf{x},\textbf{x}')$ by means of
\begin{align}
 \label{eqg1} 
\nonumber g_1(\textbf{r}) &= \int \mbox{d}^2k\, \tilde{n}(\textbf{k})\, e^{i \textbf{k}\cdot\textbf{r}} \\
 &= \int \mbox{d}^2R\,  \rho_1(\textbf{R}-\textbf{r}/2,\textbf{R}+\textbf{r}/2).
\end{align}
{A derivation of these relations is given in the supplemental material \cite{SOM}}. The function $g_1(\textbf{r})$ is a trap-averaged function, which captures the off-diagonal correlations of all particles in the system. Similarly, one can also define the central correlation function $G_{\rm 1}(\textbf{r},0) = \langle\hat{\phi}^\dagger(\textbf{r})\hat{\phi}(0)\rangle$, measured in the interference experiments \cite{Hadzibabic2006, Polkovnikov2006}, which characterizes the correlations only in the central region of the trap, where the density is approximately uniform. In general, the two functions do not contain the same information and are only equivalent in a translation invariant system \cite{SOM}. Note that due to the radial symmetry of the trapping and focusing potentials, the correlations only depend on distance and therefore it suffices to consider the azimuthally averaged function $g_1(r)$.   

\begin{figure*} [ht!]
\includegraphics [width=1\textwidth] {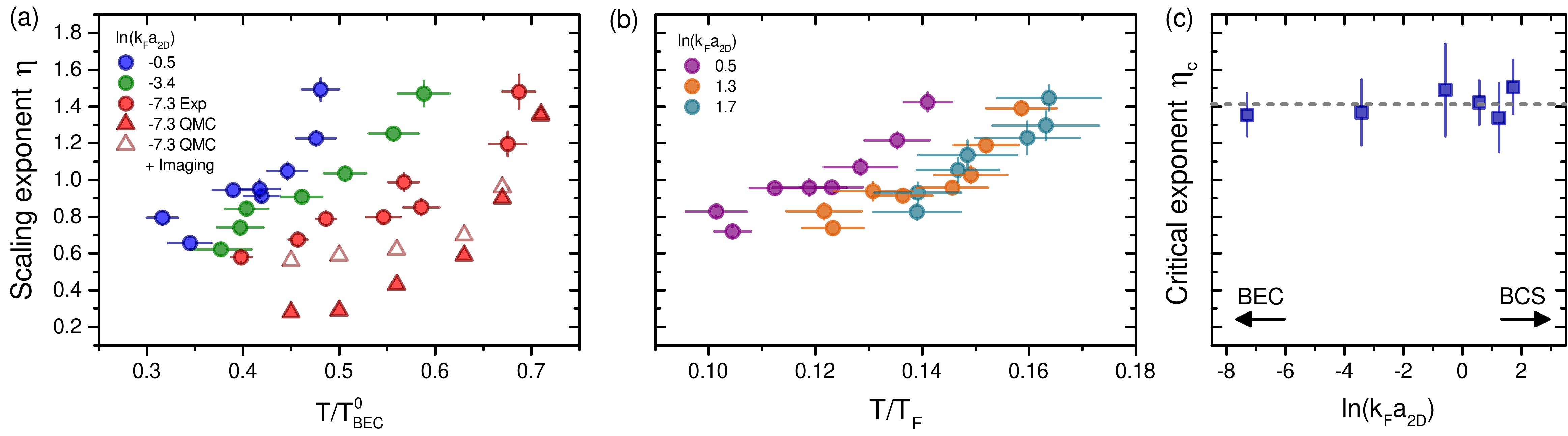}
\caption{Power-law scaling exponents across the two-dimensional BEC-BCS crossover. The temperature-dependent scaling exponent $\eta(T)$ in (a) the bosonic limit and (b) the crossover regime is shown. The relevant temperature scales in these cases are given by $T_{\rm BEC}^0$ and $T_{\rm F}$, respectively. The crossover parameter $\ln(k_{\rm F}a_{\rm 2D})$ is mildly temperature dependent. For reference we display the value at the critical temperature. For $\tilde{g}=0.60$ ($\ln(k_{\rm F}a_{\rm 2D})\simeq -7.3$) we show the prediction from QMC calculations for a Bose gas (filled red triangles) and an estimate of the effect of the finite imaging resolution present in the measured data (open red triangles) \cite{SOM}. We find an exponent which increases with temperature in agreement with BKT-theory. The power-law decay eventually ceases at $T_{\rm c}$ where a maximal exponent $\eta_{\rm c}$ is reached. (c) The value of $\eta_{\rm c}$ is approximately constant for all $\ln(k_{\rm F}a_{\rm 2D})$ where we have previously observed condensation of pairs \cite{Ries2015}. This strongly suggests that the associated phase transition are within one universality class.} 
\label{fig:etamax}
\end{figure*}
Fig.\,\ref{fig:g1r} shows the experimentally determined $g_1({r})$ for different temperatures in the strongly interacting crossover regime. The correlation functions are normalized such that $g_1(0)=1$. As expected, at high temperatures $g_1(r)$ decays exponentially with correlation lengths on the order of the thermal wavelength ($\lambda_{\rm T}  \sim 1.5\,\mu{\rm m}$). As we lower the temperature, we eventually observe the onset of coherence over an extended spatial range that corresponds to several radial oscillator lengths $\ell_r$, with $\ell_r \approx 6.8 \,\mu$m. This shows that phase fluctuations in the system are non-local and span regions of the sample where the density is not uniform. As pointed out in Refs.\,\cite{Holzmann2007, Baym2007},  such extended spatial coherence in an interacting system is a sufficient condition for superfluidity {in two-dimensional systems}.

As the temperature is lowered below a critical value, we find that the correlation function in an intermediate range $3\lambda_{\rm T} < r < 20\lambda_{\rm T}$ is well-described by a power-law decay, whereas exponential behavior is clearly disfavored. We quantify this by extracting the $\chi^2$ for both fit functions at different temperatures and observe a clear transition from exponential to algebraic decay (see Fig.\,\ref{fig:g1r} b). This qualitative change  in $g_1(r)$ provides an alternative way to determine the phase transition temperature $T_{\rm c}$ from the kink in $\chi^2(T)$ \cite{SOM}. We find that the corresponding $T_{\rm c}$ obtained in this manner agrees with the temperature associated with the onset of pair condensation that was measured in our previous work \cite{Ries2015}. 

The power-law decay of $g_1(r)$ means that the spatial coherence of the entire sample is characterized by a single exponent $\eta$. Fig.\,\ref{fig:etamax} shows the experimentally determined $\eta$ for all the interaction strengths accessed in this work.  We find $\eta(T)$ to increase with temperature until it reaches a maximal value at $T_{\rm c}$, indicating a slower fall-off of correlations at lower temperatures. Although such temperature-dependence is qualitatively consistent with BKT theory, we observe the values of the exponents to be in the range $0.6-1.4$ for the temperatures accessed in the measurement, which is substantially above the expectation of $\eta \leq 0.25$ for the homogeneous setup.

To confirm the large scaling exponents in the trapped system, we compute the one-body density matrix on the bosonic side using the QMC technique described above. This allows to determine both the trap-averaged correlation function $g_1(r)$ as well as the central correlation function $G_1(r,0)$. The trap-averaged $g_1(r)$ shows the same behavior as in the experimental case, i.e. a transition from exponential to algebraic decay at low temperatures. The corresponding QMC transition temperatures also agree with the measured values of $T_{\rm c}$ for $\tilde{g}=0.60, 1.07,$ and $2.76$. Furthermore, the maximal scaling exponent at $T_{\rm c}$ extracted from the QMC-$g_1(r)$ for $\tilde{g}=0.6$ is approximately 1.35, which is close to the experimentally determined $\eta(T_{\rm c}) \simeq 1.4$. The central correlation function $G_1(r,0)$  shows a transition to algebraic order as well -- with the same $T_{\rm c}$ as in the experiment -- but with a maximal exponent of approximately $0.25$, as expected for a homogeneous system. This finding is also in agreement with the measurement of $G_1(r,0)$ in the interference experiments \cite{Hadzibabic2006}, and is explained by the nearly uniform density in the center of the trap.

Fig.\,\ref{fig:etamax}a shows the comparison between the experimental and QMC values of $\eta(T)$ for $\tilde{g}=0.60$ ($\ln(k_{\rm F}a_{\rm 2D}) \simeq -7.3$). Although both show similar dependence on temperature, we find a considerable quantitative deviation between them. {As discussed in the supplemental material \cite{SOM}, this discrepancy can mostly be attributed to the effect of the finite imaging resolution in the measurement of $\tilde{n}(\textbf{k})$, which leads to an apparent broadening at low momenta and thus overestimates the value of $\eta$.} We show an estimate of this temperature-dependent effect on the exponents (open red triangles) in Fig.\,\ref{fig:etamax}a. There may be other effects in the experiment that contribute additionally to the deviation, such as higher order corrections to the determination of $\tilde{g}$ from the fermionic scattering parameters, and density-dependent inelastic loss processes.

\begin{figure} [ht!]
\includegraphics [width=8cm] {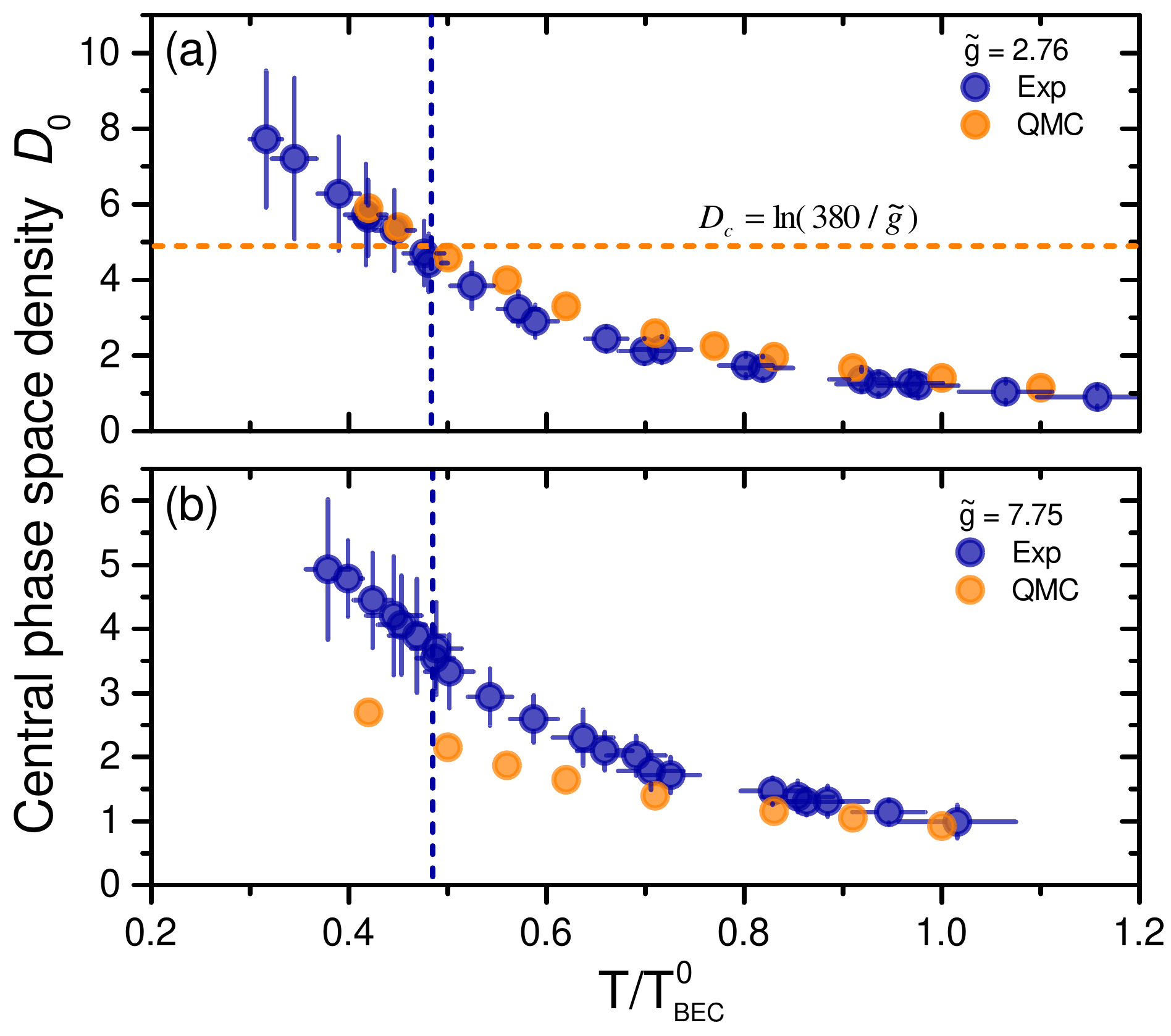}
\caption{Peak phase space density $\mathcal{D}_0=n_0\lambda_T^2$ obtained from the central density $n_0$. {The panels (a) and (b)} show experimental and simulated data (for bosons) for the coupling strengths $\tilde{g}=2.76$ and $\tilde{g}=7.75$, respectively. The vertical dashed lines indicate the corresponding critical temperatures obtained from the measured onset of algebraic order. (a) We find excellent agreement between experiment and QMC for $\tilde{g}=2.76$, providing evidence that we realize a strongly interacting 2D Bose gas. We verify the applicability of $\mathcal{D}_{\rm c} = \ln(380/\tilde{g})$ \cite{Prokofev2001} at this interaction strength (horizontal dashed line). (b) For the stronger coupling $\tilde{g}=7.75$, however, we find the bosonic simulations to deviate from the measured results, indicating fermionic superfluidity.} 
\label{fig:PSD}
\end{figure}

The experimental and simulated data raise the question why correlations in the trapped system decay with a larger scaling exponent than in the homogeneous case. To elucidate the role of inhomogeneity, we consider the bosonic field operator given by $\hat{\phi}(\textbf{r}) \simeq \sqrt{\rho(\textbf{r})}\exp(\text{i} \hat{\varphi}(\textbf{r}))$. In this representation, it is clear that one contribution to the decay of $g_1(r)$ in Eq.\,(\ref{eqg1}) comes from the spatial variation of the superfluid density $\rho(\textbf{r})$. { Using a local density approximation and assuming the superfluid density to have a Thomas--Fermi profile, we estimate a contribution of approximately $0.3 - 0.4$ \cite{SOM} to the effective exponent. Still, this fails to explain the large exponents observed in the experiment and the QMC simulations close to $T_{\rm c}$. This suggests that the increase in the effective exponents is predominantly due to phase fluctuations in the inhomogeneous system, whose  spectrum is modified by the discrete level-structure of the harmonic trapping potential and the Thomas--Fermi profile of the superfluid. This inference is further supported by calculations of phase fluctuations in a trapped 2D Bose gas at low temperatures \cite{Petrov2000B}, which indicate a trap-induced increase of the effective exponent by up to a factor of three.}

Our measurements of $g_1(r)$ and $\eta(T)$ across the two-dimensional BEC-BCS crossover provide a unique opportunity to study BKT physics even in the fermionic regime. Fig.\,\ref{fig:etamax} displays the measurement of the scaling exponent across the crossover. Remarkably, we find that -- despite varying the scattering length by several orders of magnitude -- the maximal scaling exponent $\eta_{\rm c}$ at the transition shows no dependence on the interaction strength (see Fig.\,\ref{fig:etamax}c). We note that the actual value of $\eta_{\rm c} \simeq 1.4$ might depend on parameters specific to the experiment, such as particle number and trapping frequencies. Nevertheless, the fact that $\eta_{\rm c}$ remains constant across the BEC-BCS crossover unambiguously shows that the long-range properties at the transition are independent of inter-particle interactions. This is evidence that all the observed transitions for different interaction strengths lie in the same universality class. In particular, it shows that, even as we cross over to the fermionic side ($\ln(k_{\rm F} a_{\rm 2D}) > 0$), the observed transitions are of BKT-type. 

We now turn to a quantitative investigation of local properties of the system. This allows to benchmark our measurements with (i) the QMC results for point-like bosons in the same quasi-2D trapping potential as realized in the experiment and (ii) QMC calculations of the homogenous 2D Bose gas \cite{Prokofev2001, Prokofev2002}. For this we investigate the phase space density (PSD)
\begin{align}
\label{eqPSD} \mathcal{D} = n\lambda_T^2.
\end{align}
Herein, $n$ is the {2D density }of atoms in a single hyperfine state and $\lambda_T^2=2\pi\hbar^2/M k_{\rm B}T$ is the thermal wavelength of bosons with $M$ being twice the fermion mass. Note that $n$ coincides with the density of dimers in the bosonic limit.

We first consider coupling strengths $\tilde{g}=0.60, 1.07$, and 2.76 on the bosonic side of the crossover. Fig.\,\ref{fig:PSD}a shows the comparison between the experimentally measured and QMC-computed values of the PSD in the trap-center for $\tilde{g}=2.76$. We find excellent agreement between the two data sets. In particular, at $T_{\rm c}$, the central PSD for all three $\tilde{g}$ are found to agree very well with $\mathcal{D}_{\rm c}=\ln(380/\tilde{g})$ derived for a homogeneous 2D Bose gas with weak interactions (horizontal dashed line) \cite{Prokofev2001, Prokofev2002}. This shows that the onset of algebraic correlations in the trapped system coincides with the local PSD in the center of the trap crossing the critical value of the homogeneous system \cite{Holzmann2008}. 

As we further increase $\ln(k_{\rm F}a_{\rm 2D})$, the effective boson coupling strength $\tilde{g}$ becomes very large. For $\tilde{g}=7.75$ ($\ln(k_{\rm F}a_{\rm 2D}) \simeq 0.5$),  we find substantial deviations between the experimental and QMC data for the PSD at low temperatures (see Fig.\,\ref{fig:PSD}b). Moreover, our QMC calculations show that the associated 2D Bose gas is in its normal phase for all temperatures accessed in the experiment. In contrast, the measurements show a clear superfluid phase transition at this interaction strength, as shown in Fig.\,\ref{fig:g1r}\,(lower panel). This provides evidence for the crossover to a superfluid phase whose properties are not captured by a description that assumes point-like dimers. 

Both experimental and simulated data in the bosonic limit are obtained in a highly anisotropic 3D trapping potential. Still, local observables such as the central PSD and the central correlation function $G_1(r,0)$ agree excellently in their critical properties with the theory of a homogenous 2D Bose gas and the corresponding BKT phenomena. In the case of global correlations, we showed that the inhomogeneity leads to significant deviations from the homogeneous case, most importantly an increase in the exponent of the power-law decay. However, the general features in the off-diagonal correlations -- such as the temperature-dependence of $\eta(T)$ and the independence of $\eta_{\rm c}$ from $\ln(k_{\rm F}a_{\rm 2D})$ -- suggest that the long-range physics are still captured by the ideas underlying BKT-theory for the two-dimensional XY model. 

{In conclusion, we investigated the nature of the phase transition of a trapped 2D ultracold Fermi gas. We measured for the first time the first-order correlation function of the entire system and extracted its long-range behavior. We showed that it is consistent with a description by a single power-law exponent for large distances. The transition temperature for onset of algebraic order coincides with the one obtained from the onset of pair condensation in \cite{Ries2015}. By comparing the experimental data to QMC calculations on the bosonic side, we found the system to realize a strongly interacting 2D Bose gas. The measured phase space densities and correlations on the fermionic side, instead, are not captured by a description in terms of point-like bosons, which indicates the crossover to a fermionic superfluid.}

Our measurements show that the spatial coherence even in trapped systems can be characterized by a single scaling exponent. However, understanding the underlying mechanism remains a challenge for future explorations, and may lead to a deeper understanding of phase transitions in inhomogeneous systems. \\

We thank Z. Hadzibabic, T. Enss, T. Bourdel and J. M. Pawlowski for insightful discussions. We thank T. Lompe, S. Pres and J.E. Bohn for experimental and conceptual contributions to the project. We gratefully acknowledge support from the ERC starting grant $279697$, the ERC advanced grant $290623$, the Helmholtz Alliance HA$216/$EMMI and the Heidelberg Center for Quantum Dynamics. M.H was supported by ANR-12-BS04-0022-01, ANR-13-JS01-0005-01 and DIM Nano'K from R\'{e}gion \^{I}le--de--France. M.G.R. and I.B. acknowledge support by the Landesgraduiertenf\"orderung Baden-W\"urttemberg. 

P.A.M. and I.B. contributed equally to this work.

\bibliography{2Dpaper_ver02}

\cleardoublepage

\section*{\large SUPPLEMENTAL MATERIAL}
\setcounter{figure}{0}
\renewcommand{\figurename}{Fig.\,S}

\section{Extracting the first-order correlation function}

From the matterwave focussing technique we obtain the momentum distribution
\begin{align}
 \tilde{n}(\textbf{k}) = \langle \hat{a}^\dagger_{\textbf{k}} \hat{a}_{\textbf{k}}\rangle
\end{align}
of the trapped gas. Herein, $\hat{a}^\dagger_{\textbf{k}}$ is the creation operator for a particle with wave vector $\textbf{k}$. Note that $\tilde{n}(\textbf{k})$ involves an average over all particles in the trap. 

We now show that the Fourier transform of this function coincides with $g_1(\textbf{r})$ in Eq. (1). For this we recall that $\hat{a}_{\textbf{k}}$ is obtained from $\hat{\phi}(\textbf{x})$ by means of
\begin{align}
 \hat{\phi}(\textbf{x}) = \int \frac{\mbox{d}^2k}{(2\pi)^2}e^{\rmi \textbf{k}\cdot\textbf{x}} \hat{a}_{\textbf{k}}.
\end{align}
Inserting this definition into (1) with the substitution $\textbf{s}=\textbf{R}+\textbf{r}/2$ we arrive at
\begin{align}
 \nonumber g_1(\textbf{r}) &= \int \mbox{d}^2s \ \rho_1(\textbf{s},\textbf{r}+\textbf{s})\\
 \nonumber &= \int \mbox{d}^2s \ \langle \hat{\phi}^\dagger(\textbf{s})\hat{\phi}(\textbf{r}+\textbf{s}) \rangle\\\
 \nonumber &= \int \mbox{d}^2 s \frac{\mbox{d}^2k}{(2\pi)^2} \frac{\mbox{d}^2k'}{(2\pi)^2} \ e^{-\rmi (\textbf{k}-\textbf{k}')\cdot\textbf{s}} e^{\rmi \textbf{k}'\cdot\textbf{r}} \langle \hat{a}^\dagger_{\textbf{k}}\hat{a}_{\textbf{k}'}\rangle\\
 \nonumber &= \int \frac{\mbox{d}^2k}{(2\pi)^2} \frac{\mbox{d}^2k'}{(2\pi)^2}  (2\pi)^2 \delta^{(2)}(\textbf{k}-\textbf{k}') e^{\rmi \textbf{k}'\cdot\textbf{r}} \langle \hat{a}^\dagger_{\textbf{k}}\hat{a}_{\textbf{k}'}\rangle\\
 &= \int \frac{\mbox{d}^2k}{(2\pi)^2}e^{\rmi \textbf{k}\cdot\textbf{r}}\langle \hat{a}^\dagger_{\textbf{k}}\hat{a}_{\textbf{k}}\rangle.
\end{align}
The expression in the last line is the Fourier transform of $\tilde{n}(\textbf{k})$. 

In a translation invariant situation, the one-body density matrix can be written as $\rho_1(\textbf{x},\textbf{x}')=f(\textbf{x}-\textbf{x}')$, with some function $f$. In this case we have $g_1(\textbf{r}) = \int \mbox{d}^2s \ \rho_1(0,\textbf{r}) \propto G_1(0,\textbf{r})$. The difference between the trap-averaged correlation function, $g_1(\textbf{r})$, and the central correlation function, $G_1(0,\textbf{r})$, then consists of an overall (volume) factor, which vanishes in the normalization procedure.

\begin{figure} [hb!]
\includegraphics [width=7.5cm] {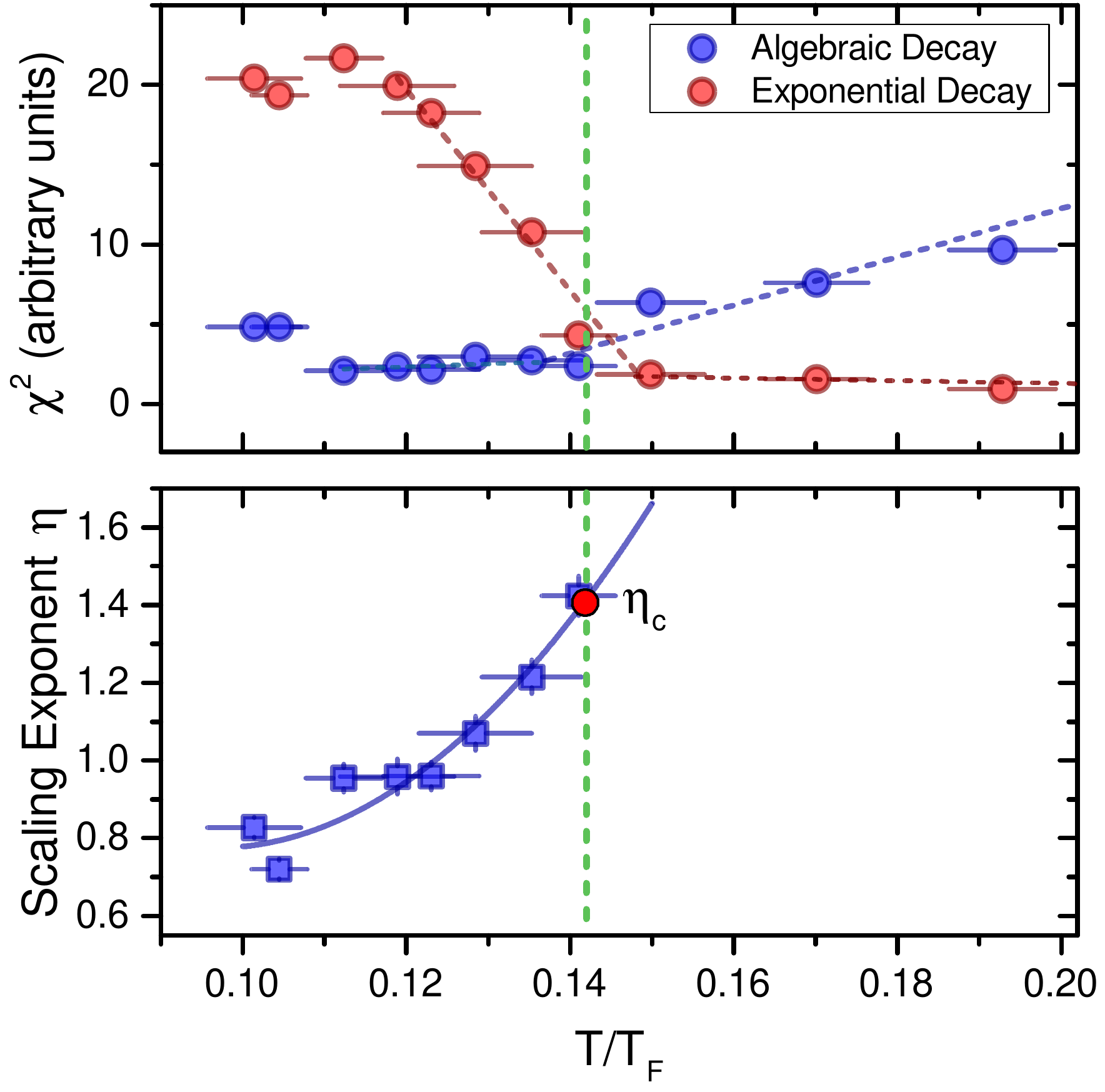}
\caption{Extracting critical temperature $T_{\rm c}$ and critical exponent $\eta_{\rm c}$ at 812 G. The upper panel shows the $\chi^2$-values for exponential (red) and power law (blue) fits of $g_1(r)$, respectively. Lower values of $\chi^2$ indicate a better description of the data. This allows to determine the critical temperature as the temperature of onset of algebraic decay in $g_1(r)$. The lower panel shows the scaling exponent $\eta(T)$ below $T_{\rm c}$. We determine the critical exponent (red circle) by extrapolating $\eta(T)$ using a polynomial fit.} 
\label{fig:rsqr}
\end{figure}

\textbf{Momentum resolution and coherence length:} The matterwave focusing lens used for the measurement of $\tilde{n}(\textbf{k})$ has a magnification factor $M\omega_{\rm lens}$, where $\omega_{\rm lens} = 2\pi \times 10\,$Hz is the trap frequency of the focusing potential and $M$ is the molecular mass. From this, the effective momentum space resolution can be obtained according to $\Delta k = M\omega_{\rm lens} \Delta x / \hbar \simeq 0.035\,\mu{\rm m}^{-1}$, with $\Delta x \sim 5\,\mu{\rm m}$ being the spatial imaging resolution. This means that the largest coherence length that is accessible in $g_1(r)$ is approximately $L = 2\pi/\Delta k \sim 105\,\mu{\rm m}$.

\section{Extracting critical temperature and critical scaling exponent}

\begin{table*} [htb!]
\begin{center}
\small
  \begin{tabular}{|c| c | c | c | c |c | c |} \hline
    \: $B$ [Gauss] \: &\:  $\tilde{g}$  \:&\: $\ln(k_Fa_{2D})_{T_c}$ \:& \: $T_{\rm c}/T^0_{\rm BEC}$ (stat.)(sys.)\:&\:$T_{\rm c}/T_{\rm F}$ (stat.)(sys.)\:&\: $T_{\rm c}/T_{\rm F}$ (stat.)(sys.) \:&\:$\eta_{\rm c}$ (sys.)\:   \\ 

	& &(stat.)(sys.) &(algebraic decay)& (algebraic decay) & (pair condensation \cite{Ries2015}) & \\    
    
    \hline 
    & & & & & & \\
    692	& \,\,0.59 	&-\,7.30 \,(4) $\left(_{-5}^{+4}\right)$ & 0.695 (25)\,$\left(_{-87}^{+103}\right)$		&\,0.091\,(21)  $\left(_{-13}^{+16}\right)$			& 0.089 \, (15)  $\left(_{-13}^{+14}\right)$	& 1.35\,\,(12)  	\\ [5pt]
 
	732	& \,\,1.07 	&-\,3.42 \,(2) $\left(_{-6}^{+4}\right)$ & 0.590 (24)\,$\left(_{-74}^{+105}\right)$		&\,0.094\,(19)  $\left(_{-14}^{+19}\right)$			& 0.100 \, (22)  $\left(_{-15}^{+17}\right)$	& 1.36\,\,(18)   	\\ [5pt]
	
    782	& \,\,2.76	&-\,0.59 \,(1) $\left(_{-7}^{+4}\right)$ & 0.483 (20)\,$\left(_{-59}^{+80}\right)$	&\,\,\,0.114 \,(19)  $\left(_{-16}^{+23}\right)$		& 0.129 \, (35)  $\left(_{-18}^{+24}\right)$	& 1.48\,\,(25)   \\ [5pt]
    
	812	& \,\,7.75	&\,\,\,0.57 \,(1) $\left(_{-7}^{+2}\right)$	& 0.485 (22)\,$\left(_{-58}^{+73}\right)$ &\,\,\,0.142 \,(24)  $\left(_{-18}^{+28}\right)$		& 0.146 \, (25)  $\left(_{-23}^{+50}\right)$	& 1.42\,\,(12)   \\ [5pt]
	
	832	& \,\,-		&\,\,\,1.23 \,(1) $\left(_{-8}^{+2}\right)$	& - &\,\,\,0.157 \,(30)  $\left( _{-20}^{+33}\right)$		& 0.167 \, (39)  $\left( _{-34}^{+48}\right)$	& 1.33\,\,(18)   \\ [5pt]
	
	852	& \,\,-	&\,\,\,1.72 \,(1) $\left(_{-9}^{+2}\right)$ & - &\,\,\,0.166 \,(40)  $\left( _{-22}^{+42}\right)$		& 0.167 \, (27)  $\left( _{-20}^{+37}\right)$	& 1.50\,\,(15)   \\ [5pt]
  \hline 
  \end{tabular}
\caption{We show the data for the measured critical temperatures and critical exponents for all interaction strengths considered in this work. For each magnetic field value we display the corresponding bosonic coupling strength $\tilde{g}$ and the crossover parameter $\ln(k_{\rm F}a_{\rm 2D})$ at the critical temperature. The Fermi momentum is defined as $k_{\rm F}=\sqrt{4\pi n}$, where $n$ is the central density of atoms in a single hyperfine state. The critical temperature obtained from the onset of algebraic decay is shown with respect to $T_{\rm BEC}^0$ and $T_{\rm F}$. For comparison, we also list the transition temperatures for the onset of pair condensation determined in \cite{Ries2015}. The scaling exponent $\eta_{\rm c}=\eta(T_{\rm c})$ is obtained from the extrapolation described in Sec. II.}
  \label{tab:tc}
\end{center}
\vskip -0.5cm
\end {table*}

The qualitative change of the decay of correlations when lowering the temperature allows for a determination of the critical temperature $T_{\rm c}$ for each value of $a_{\rm 2D}$. For this purpose we fit both an exponential ($g_1(r)=a e^{-r/\xi}$) and algebraic ($g_1(r)=ar^{-\eta}$) model function to the intermediate length scales of $g_1(r)$ and extract the associated $\chi^2$-value. A smaller value of $\chi^2$ corresponds to a better fit. In Fig.\,S\,\ref{fig:rsqr} (upper panel) we display $\chi^2(T)$ for 812\,G ($\tilde{g}=7.75$). A sharp transition in the behavior of correlations is visible at a certain temperature, which we associate with the critical temperature $T_{\rm c}$. To determine $T_{\rm c}$ we piecewise linearly interpolate $\chi^2(T)$ according to

\begin{align}
 \chi^2_{\rm alg}(T) &= c_1 \theta(T_{\rm c}^{(1)}-T)+c_2(T-T_{\rm c}^{(1)})\theta(T-T_{\rm c}^{(1)}),\\
 \chi^2_{\rm exp}(T) &= c_3(T_{\rm c}^{(2)}-T)\theta(T_{\rm c}^{(2)}-T)+c_4\theta(T-T_{\rm c}^{(2)})
\end{align}
for the algebraic (alg) and exponential (exp) fits, respectively. Here $\theta(x)$ is the Heaviside step function. We generically find $c_1$ and $c_4$ to be small, which justifies the choice of a power law fit at low temperatures, and an exponential fit at large temperatures. Furthermore, the temperatures $T_{\rm c}^{(1)}$ and $T_{\rm c}^{(2)}$ coincide within a few percent. We set $T_{\rm c}=(T_{\rm c}^{(1)}+T_{\rm c}^{(2)})/2$ to obtain the transition temperature.

Using the critical temperature $T_{\rm c}$ found in this manner, we extract the scaling exponent $\eta_{\rm c}$ at the transition. For this purpose we extrapolate the experimental data points for $\eta(T)$ by means of a quadratic polynomial fit, see Fig.\,S\,\ref{fig:rsqr} (lower panel) and extract the value of $\eta(T_{\rm c})$. We list the measured critical temperatures and scaling exponents in Table \ref{tab:tc}.

\textbf{Errors:} The method described above to extract the critical exponents contains some uncertainties. 
The statistical errors on the measured exponents are quite small and the error on $\eta_{\rm c}$ as shown in Fig.\,2.c is mainly due to the uncertainty in the extrapolation of $\eta(T)$. The error bars are obtained according to $\delta \eta_{\rm c} = |\eta(T_{\rm c}^{(1)}) - \eta(T_{\rm c}^{(2)})|/2$.  

\section{Systematic Effects}
The errors on experimental quantities shown in the main text are statistical uncertainties of our measurements. The systematic effects in our measurements have been discussed in detail in \cite{Ries2015}. In this work, we additionally introduce the temperature scale $T/T_{\rm BEC}^0$ and extract the scaling exponent $\eta(T)$ from $g_1(r)$. These quantities are systematically affected by the following factors:

\subsection*{Uncertainty in particle number} 
We determine the number of atoms in the cloud from in-situ absorption images. $N$ is affected by the intensity of the imaging beam, magnification of the imaging system and the small population of atoms in the non-central pancakes of the trapping potential. The ideal gas condensation temperature depends on the atom number according to $T_{\rm BEC}^0 = \sqrt{6N} \frac{\hbar \omega_{r}}{\pi k_{\rm B}}$, and hence it is affected by the uncertainty in $N$. In addition to experimental uncertainties, the measurements also contain atom number fluctuations of about 10-15$\%$.

\subsection*{Imaging effects}
The correlation function $g_1(r)$ is determined by means of a 2D Fourier transform of the in-plane pair momentum distribution $\tilde{n}(\textbf{k})$. The momentum distribution is obtained using a matterwave focusing technique which consists of a ballistic expansion of the gas in a harmonic potential for a quarter of the trap period ($\tau/4 = 25\,\text{ms}$) and subsequently imaging the planar density distribution. 

\begin{figure} [hb!]
\includegraphics [width=8cm] {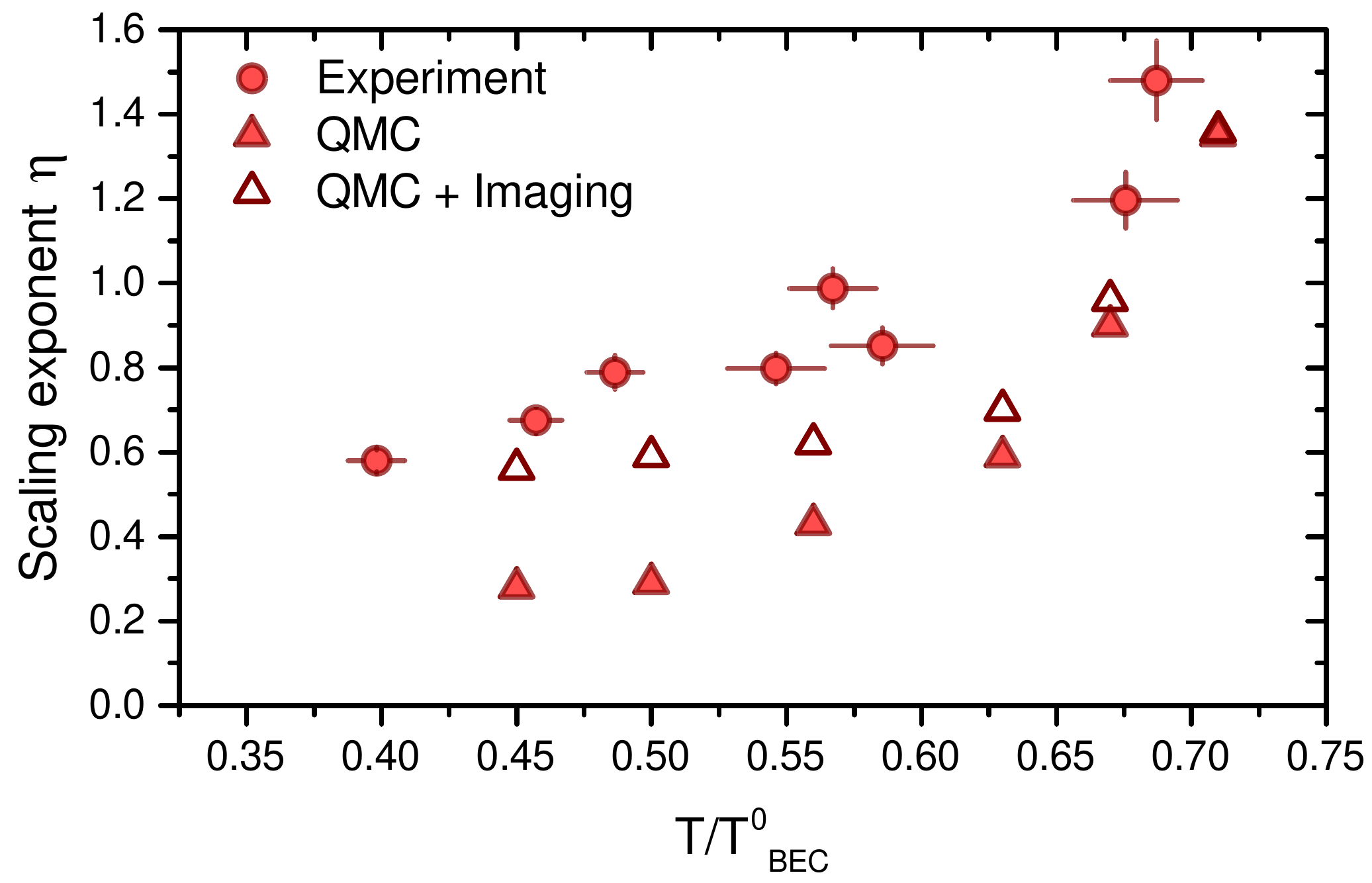}
\caption{Effect of finite imaging resolution on the scaling exponent $\eta$ for $\tilde{g}=0.60$. The imaging simulations are performed assuming the $g_1(r)$ and the corresponding exponents $\eta$ obtained from QMC computations (filled triangles). The open triangles show the exponent extracted after the simulated imaging. We find a substantial temperature-dependent deviation that is qualitatively consistent with the experimentally measured exponents (filled circles).} 
\label{fig:imaging}
\end{figure}
\newpage

The onset of algebraic decay at large distances in $g_1(r)$ corresponds to a peak at low momenta in the measured $\tilde{n}(\textbf{k})$. Naturally, a peakier $\tilde{n}(\textbf{k})$ leads to a broader $g_1(r)$ and hence a smaller scaling exponent. This method of determining $g_1(r)$ and $\eta(T)$ from the momentum distribution is limited by two main factors:

\textbf{a. Vertical expansion during TOF:} As shown in \cite{Murthy2014}, the gas expands rapidly in the vertical direction upon release from the trapping potential. After 25\,ms, the vertical extent of the cloud is approximately 500\,$\mu$m. This can lead to some parts of the sample exceeding the depth of focus of the imaging system, which in turn causes some distortion in the measured density distribution $n(\textbf{r},\tau/4)$. In general, it leads to a broadening of the $\tilde{n}(\textbf{k})$ which has the effect of increasing the measured scaling exponent $\eta$.

\textbf{b. Finite imaging resolution:} 
As we lower the temperature, the momentum distribution becomes narrower. The measured $\tilde{n}(\textbf{k})$ is the convolution of the actual momentum distribution with the finite resolution of our imaging setup. This convolution leads to a broadening of the momentum distribution for small $k$ and hence a steeper decay of $g_1(r)$ for large $r$. Moreover, this broadening effect is larger for distributions that  are closer in width to the resolution limit. Intuitively, this means that the distortion caused by the imaging resolution is enhanced at lower temperatures. For the extracted scaling exponents, this leads to a temperature-dependent deviation between the actual value and the measured value. The scaling exponent is always overestimated due to the finite imaging resolution.

To estimate the contributions of the vertical extent of the sample and the finite imaging resolution on the extracted scaling exponents, we perform a simulation of the imaging setup which consists of two lenses ($f=80\,$mm) in a 4f-configuration and a CCD camera. 

From the QMC-computed $g_1(\textbf{r})$ at each temperature, we determine $\tilde{n}(\textbf{k})$ by means of an inverse Fourier transform. This corresponds to a 2D spatial distribution $\tilde{n}(\textbf{r})$ of the cloud with $|\textbf{r}| = \hbar|\textbf{k}|/M\omega_{\rm lens}$, where $\omega_{\rm lens}$ is the trap frequency of the focusing potential. To account for the vertical extent of the cloud after time-of-flight, we construct a 3D density distribution according to $\tilde{n}(x, y, z) = L_{\rm z}^{-1} \cdot \tilde{n}(x, y)$ for $ -L_{\rm z}/2 < z < L_{\rm z}/2$, where $L_{\rm z} \approx 500\,\mu$m is the vertical size of the cloud. While this is not truly reflective of the actual distribution in the experiment, it is sufficient to capture the essential effect of the vertical size of the gas.

We consider the imaging effect of a thin section of the cloud defocused by a distance $z$ from the focal plane at $z=0$. Using the paraxial wave equations, we compute the propagation of the imaging beam, approximated as a plane wave, through this section and through the lenses. The resulting intensity distribution contains the effect of the defocus as well as the finite resolution of the lenses. We perform this computation for all $-L_{\rm z}/2 < z < L_{\rm z}/2$ and integrate the resulting intensity distributions in the 4f plane according to $I_{\rm 4f}(x,y) = \int_{-L_z/2}^{L_z/2} I(x,y,z)\mathrm{d}z$. This yields the imaged column density $n'(x,y)$ and hence the imaged momentum distribution $\tilde{n}'(\textbf{k})$. From this we obtain the imaged $g_1'(r)$ and the corresponding scaling exponent $\eta'(T)$. Fig.\,S\,\ref{fig:imaging} shows the comparison of exponents extracted from the experimental measurement, QMC computations and the imaging simulations, for typical simulation parameters. 
 
From the imaging simulations, it is clear that the finite imaging resolution causes a significant overestimation of the scaling exponent. We find that the main contribution to the deviation is in fact from the finite resolution in the radial plane and the effect of the vertical extension of the cloud is mild. This explains the strong temperature-dependence of the discrepancy between measured and QMC exponents as shown in Fig.\,S\,\ref{fig:imaging}.

\section{Local density approximation}

We estimate the influence of the Thomas--Fermi (TF) profile of the superfluid density on the correlations within a local density approximation. The result of the analysis can be summarized in the following statements: (1) The TF profile results in an additive contribution $\eta_{\rm TF}$ to the scaling exponent, which is approximately 0.3 at low temperatures; (2) The temperature-dependence of $\eta_{\rm TF}$ is mild and from the QMC data for the density profiles we estimate the value of $\eta_{\rm TF}$ at the transition to be approximately 0.4. We conclude that the large scaling exponents observed in the experiment, in particular the value $\eta_{\rm c} \simeq 1.4$ at the transition, are mostly due to phase fluctuations in the inhomogeneous sample. 

We assume that, within a phase-amplitude representation, the bosonic field in the superfluid phase can be written as
\begin{equation}
\hat{\phi}(\textbf{r}) = \sqrt{\rho(\textbf{r})} \exp i\hat{\varphi}(\textbf{r}),
\end{equation}
where $\hat{\phi}$ is an operator, but $\rho(\textbf{r})$ is a function. We approximate $
\rho(\textbf{r})$ to be given by a TF profile according to $\rho(\textbf{r}) \simeq \rho_0 (1- r^2/R^2_{\rm TF})\theta (1- r^2/R^2_{\rm TF})$, where $\theta(x)$ is the Heaviside step function, and $R_{\rm TF} = (2\hbar^2\tilde{g}n_0)^{1/2}/M\omega_r$ is the radius of the superfluid core. The particular shape of $\rho(\textbf{r})$, however, is not essential for the following conclusion. We then find
\begin{equation}
g_1(\textbf{r}) = \int \mbox{d}^2s \sqrt{\rho(\textbf{r})\rho(\textbf{s} + \textbf{r})}\langle e^{i(\hat{\varphi}_{\textbf{s}} - \hat{\varphi}_{\textbf{s} + \textbf{r}})} \rangle
\end{equation}
for the trap-averaged correlation function. \\

\begin{figure} [ht!]
\includegraphics [width=7cm] {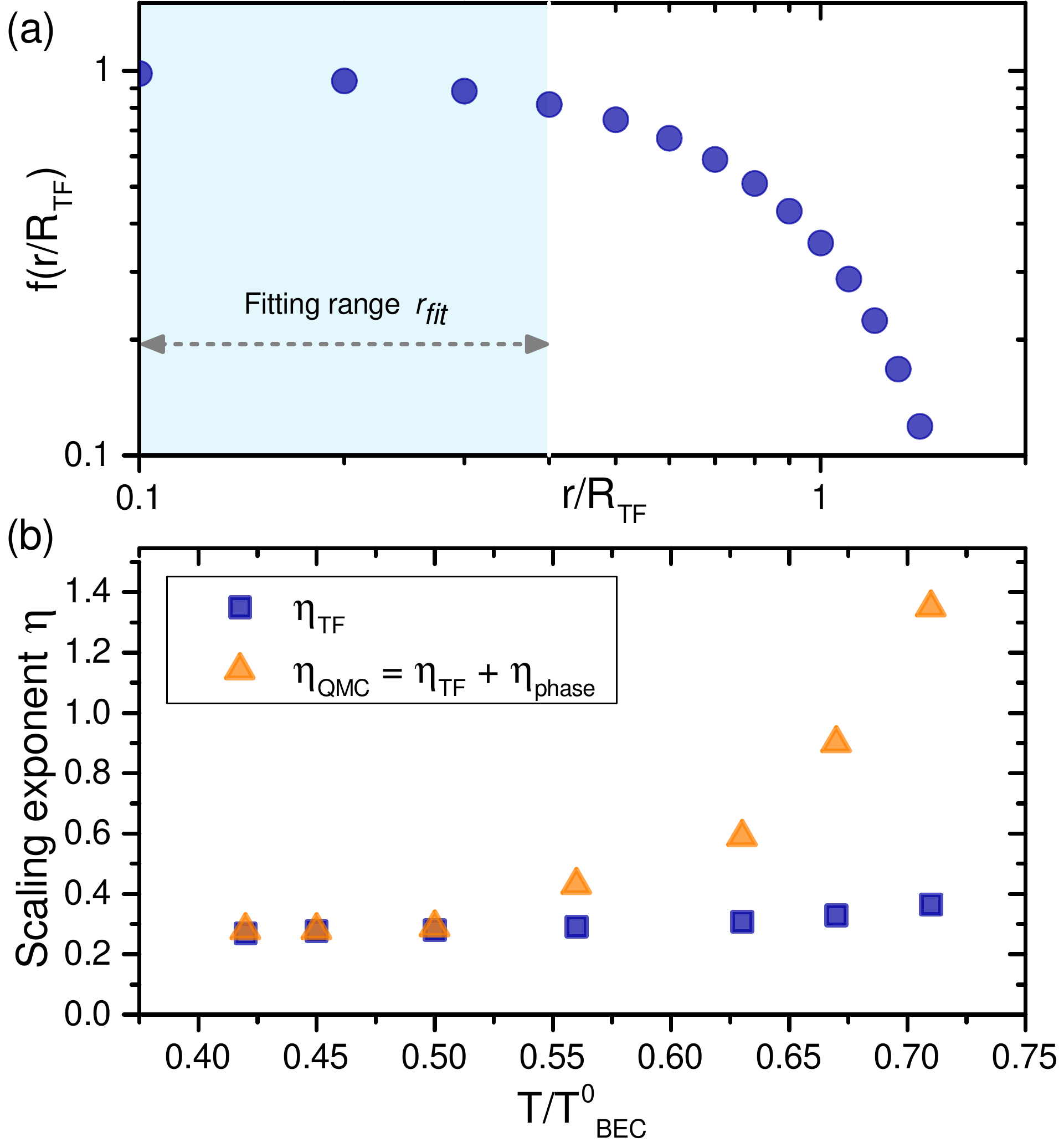}
\caption{Estimated effect of the Thomas--Fermi (TF) profile onto the decay of correlations in a local density approximation. Panel (a) shows the function $f(r/R_{\rm TF})$ from Eq.\,(\ref{Eqfr}) which multiplies the algebraic decay due to phase fluctuations. Our fitting range, highlighted by the blue shaded region, is given by $r/R_{\rm TF}\lesssim 0.4$, where the function is rather flat. The fall-off of $f(r/T_{\rm TF})$ leads to an additive contribution $\eta_{\rm TF}$ to the extracted scaling exponent. The latter is shown in panel (b). At low temperatures, where phase fluctuations are small, we have $\eta \simeq \eta_{\rm TF} \simeq  0.3$. Even at large temperatures, we only have $\eta_{\rm TF}\simeq 0.4$. This effect cannot fully explain the large exponents found from the QMC and experimental data. Hence the associated decay of correlations must be mostly due to phase fluctuations.} 
\label{fig:LDA}
\end{figure}

Approximating the phase fluctuations to be translation invariant we write $\langle e^{i(\hat{\varphi}_{\textbf{s}} - \hat{\varphi}_{\textbf{s} + \textbf{r}})} \rangle \sim |\textbf{r} - \textbf{r}'|^{-\eta_{\rm phase}(T)}$ with a temperature-dependent exponent $\eta_{\rm phase}(T)$, which is assumed to be constant throughout the superfluid region. We arrive at
\begin{equation}
g_1(\textbf{r})  \sim r^{-\eta_{\rm phase}(T)}f(\frac{r}{R_{\rm TF}})
\label{Eqfr}
\end{equation}
with the function $f$ being shown in Fig.\,S\,\ref{fig:LDA}a.

The spatial decay of the function $f$ results in an additive contribution $\delta \eta_{\rm eff} = -\mbox{d}\ln f / \mbox{d} \ln(r/R_{\rm TF}) \simeq 1.97(r/R_{\rm TF})^2 $ to the measured scaling exponent. In our experiment, we have $R_{\rm TF} \simeq 100\,\mu\text{m}\sqrt{\tilde{g}}$ and typical fitting ranges are $r \leq r_{\rm fit} \simeq 4\ell_r = 30\,\mu\text{m}$. Defining the TF contribution as $\eta_{\rm TF} = \delta \eta_{\rm eff}(r_{\rm fit}/R_{\rm TF})$ we have
\begin{equation}
\eta(T) = \eta_{\rm TF}(T) + \eta_{\rm phase}(T)
\end{equation}
for the total scaling exponent extracted from the data. The temperature dependence of the TF contribution results from the temperature dependence of the central density $n_0(T)$. We use the latter from the QMC data for the density profiles to compute $\eta_{\rm TF}(T) = \eta_{\rm TF}(T_0)\frac{n_0(T_0)}{n_0(T)}$, where $T_0$ is a reference temperature. We choose the latter to be small. The estimated TF contribution $\eta_{\rm TF}(T)$ for $\tilde{g}=0.60$ is shown in Fig.\,S\,\ref{fig:LDA}b.

\end{document}